\documentclass[english]{article}
\usepackage[T1]{fontenc}
\usepackage[latin1]{inputenc}

\makeatletter

\usepackage{babel}
\makeatother
\begin{document}

\title{Defining a Comprehensive Threat Model for High Performance Computational
Clusters}

\author{Dmitry Mogilevsky$^{*\ddagger}$, Adam Lee$^{\ddagger}$, William
Yurcik$^{*}$\\
\\
$^{*}$National Center for Supercomputing Applications (NCSA)\\
\\
 $^{\ddagger}$Computer Science Department \\
University of Illinois at Urbana Champaign (UIUC)\\
\\
 \{dmogilev,adamlee,byurcik\}@ncsa.uiuc.edu}

\maketitle
\begin{abstract}
Over the past decade, high performance computational (HPC) clusters
have become mainstream in academic and industrial settings as accessible
means of computation. Throughout their proliferation, HPC security
has been a secondary concern to performance. It is evident, however,
that ensuring HPC security presents different challenges than the
ones faced when dealing with traditional networks. To design suitable
security measures for high performance computing, it is necessary
to first realize the threats faced by such an environment. This task
can be accomplished by the means of constructing a comprehensive threat
model. To our knowledge, no such threat model exists with regards
to Cluster Computing. In this paper, we explore the unique challenges
of securing HPCs and propose a threat model based on the classical
Confidentiality, Integrity and Availability security principles. 
\end{abstract}

\section{Introduction\label{sec:Introduction}}

Cluster computing now constitutes over 60\% of the top 500 high performance
computing resources in the world\cite{top500}, with top-performing
clusters such as IBM's Blue Gene reaching peak speed of over 180 TFlops
across more than 65,000 nodes \cite{top500-ibm}. HPCs are used for
a variety of research and industry tasks many of which are either
mission-critical or sensitive by nature, making clusters an attractive
target for industry espionage or sabotage. Additionally, the cluster
itself, with its highly desirable resources such as powerful computational
capacities, high-bandwidth network connection and massive storage,
which can be employed for a DoS attack, brute force password cracking\cite{perrine-teracrack}
or illegitimate FTP servers, make it an attractive target for attackers.
An example of such an attack took place in spring of 2004, when attacks
on facilities in multiple institutions took place \cite{krebs}.

Attempts to secure the cluster computing environments currently suffer
from lack of an integrated security approach that takes advantage
of the intrinsic properties of cluster computing. Applying traditional
security measures to individual nodes of the cluster is an inadequate
measure, as it fails to take into consideration the overall context
of the system. For example, a node may be communicating on a port
that appears legitimate to the security system running on that node.
If the security system were able to interface with the scheduler,
it would learn that no job is scheduled to be running on that node,
therefore there should be no communication taking place\cite{Yurcik1-EmergentProperties}.
In order to be able to define a comprehensive approach to securing
a cluster, we must first strive to completely understand the threats
and security risks that are present in cluster computing environment. 

The best approach to analyzing the threats faced by a computing environement
is through a systematic threat-analysis approach, or threat modelling.
Threat modelling involves systematically identifying the assets in
the system, creating an architectural overview of the system, and
identifying the threats at each stage of the system\cite{microsoft-threadmodel}.
Once the threats are identified, a risk assesment analysis is performed
to determine whether it is more efficient to mitigate the threat or
accept the risk of it being exploited. Security mechanisms are then
developed to mitigate the threats which are determined to be unacceptable.
This process allows security engineers to effectively identify which
security measures are necessary. This ensures that the necessary mechanisms
are put in place, while the unnecessary ones are left unimplemented\cite{rahib}.

The purpose of this paper is to effectively and comprehensively identify
the threats faced by cluster computing%
\footnote{The risk assessment portion of the security analysis process depends
in large part on the workload and data sensitivity of each individual
cluster, as well as details of its architecture, and as such, best
left for the cluster administrator to perform. %
}. We chose to use the classical Confidentiality, Integrity and Availability
(CIA) security model, a well-accepted and time-proven paradigm in
the security community, as a basis for our threat model. We discuss
the unique aspects of a clustered computing network as compared to
traditional networks, and how the challenges these aspects present
within the framework of the CIA model. To the best of our knowledge,
this is the first attempt to use a structured CIA approach to creating
a comprehensive threat model for clusters.

The remainder of the paper is organized as follows. In Section \ref{sec:Security-Challenges-of}
we discuss the unique aspects of cluster computing as they relate
to security assurance. In Section \ref{sec:Previous-Work} we briefly
survey the previous work on defining a comprehensive cluster security
model. We then discuss our threat model in section \ref{sec:Threats-facing-Cluster}
of the paper, and offer a summary and some concluding remarks in section
\ref{sec:Summary-and-Conclusions}.

\section{Security Challenges of Cluster Computing\label{sec:Security-Challenges-of}}

As high speed computing continues to shift from mainframe to commodity
CPU clusters, it is important to note several emerging properties
in this new computing environment, which directly contribute to the
difficulty in maintaining a secure computing environment.

\subsection{Clusters are highly customizable}

Clusters can be thought of as high speed networks of commodity processors.
As such, there is not a single definition of what a typical cluster
'profile' is. Varied factors among clusters include node quantity,
CPU and chipset choice, operating system, cluster management software,
such as Rocks and OSCAR, and interconnect, such as Infiniband, Myrinet
and many others\cite{baker2}.

\subsection{Clusters are often highly heterogeneous}

Frequently, clusters are deployed incrementally, and with different
components. As such, clusters can often be found to contain a highly
heterogenous mix of hardware devices and software configurations.
This presents a security challenge that is both local and distributed.
On one hand, each distinct configuration has unique needs in terms
of patching and node hardening. On the other hand, lack of homogeneity
hinders deployment of integrated security solution across the whole
cluster \cite{Pourzandi-"clustersec"}.

\subsection{Performance-First mentality}

Clusters are designed to be a high performance computing tool. As
such, it is a common practice to maximize the performance potential
and accessibility of clusters, often at expense of security. One such
example is the practice of exposing all of the cluster's nodes to
public networks in order to allow users to login directly to computing
nodes and run their jobs. The emergence of cluster grid computing
and on-demand computing has accenuated this problem, since these paradigms
require all nodes to be accessible to outside connections for optimal
effectiveness. Another example is disabling SSH authentication between
nodes to reduce job start up time.

\subsection{Edge-based security measures model}

As part of the performance-first paradigm described above, security
measures in clusters, such as secure authentication login are often
concentrated along the edges of the cluster. As a result, once an
attacker gains access to the cluster, there may not be significant
obstacles to what he can do inside.

\subsection{Cluster size}

As clusters grow in size, it become increasingly difficult to guarantee
secure state operation in every one of the components. Increasing
the size of the cluster also increases the attack space and heightens
the possibility that a deteremined attacker can locate and exploit
a possibility in the system to gain access to one of the nodes. Given
the security oversight discussed in previous section, finding one
vulnerable node is often all an attacker needs.

These aspects clearly differentiate clusters from classical network
security realms. As a result, cluster security cannot be addressed
in the same way that tradional network security is addressed. In light
of the fact that a cluster presents a very different set of challenges
to both attackers and security engineers, and contains vastly different
set of assets to potential attackers, cluster security must be examined
within the framework of a different threat model than one faced by
traditional networks.

\section{Previous Work\label{sec:Previous-Work}}

To the best of our knowledge, no work has been done on developing
a comprehensive threat model for cluster computing, to the best of
our knowledge. In \cite{naqvi}, the authors define a threat model
for Grid computing at a very high level and without focusing on the
specific security challenges of the clusters that are outlined in
this work. In \cite{torres}, three specific attacks on clusters are
presented. In \cite{Pourzandi-"clustersec",Yurcik1-EmergentProperties},
the unique properties of cluster security are addressed, and a limited
threat model, along with a proposal for an integrated cluster security
tool is given.

\section{Attack Identification\label{sec:Threats-facing-Cluster}}

To create a strong threat model of a system, several questions must
be addressed. The first step of developing a threat model is to identify
the potential assets of the system. An asset in threat modeling is
defined as an entity or a feature of a system that is of interest
to the attacker, and as such, gaining access to these assets is the
motivation behind the attack \cite{microsoft}. It is the case, however,
that assets represent entities or features that are desired for legitimate
users as well, therefore it is not a practical approach to eliminate
the motivation for attacks by eliminating the assets. Secondly, we
must identify potential entry points into the system which can be
exploited by the attacker in order to gain illegitimate access. Entry
points can be intentional, such as a public login script, or unintentinal,
such as an open port or a buffer overflow vulnerability in a running
library. Finally, given the existing assets and entry points, we enumerate
what attacks can be launched in order to gain access to each given
asset.

It is imperative to follow the above steps when designing a threat
model. Threat modeling requires a systematic and repeatable approach,
which is not achievable by simply brainstorming the question {}``How
can I be attacked?'' \cite{suvda}. One must consider not only the
assets at risk and the vulnerabilities of the system, but also non-technical
questions, such as, who is going to be launching the attacks (defacers,
industrial spies, script kiddies, etc.), and what is the motivation
behind the attacks (financial gain, access to computing resources,
etc) \cite{rahib}. 

A few words on our attacker model. We refer to anyone who wishes to
circumvent the normal operation state of the cluster as an 'attacker'.
This is a diverse qualification that can include a bored teenager
hacking from home, a malicious hacker attempting to steal confidential
information from the cluster, a disgrunted employee, or a legitimate,
but dishonest cluster user who is attempting to get more cluster resources
at the expense of other users. Though these people may have very different
goals in mind, we group them together as potential attackers.

We choose to classify attacks on clusters using the Confidentiality-Integrity-Availability
threat classification. CIA security model is a seminal classification
model in information assurance studies and provides a more clear-cut
separation than newer, alternative models such as STRIDE \cite{microsoft}.

\textbf{Assets.} As we have briefly mentioned already, a large-scale
cluster is highly attractive to attackers both for the data contained
in it and the physical resources that it provides. We identify the
following assets in the cluster that an attacker might try to get
access to.

\begin{itemize}
\item User login data
\item User job data
\item System logs
\item Scheduler
\item Storage systems
\item Intranode network fabric
\item Computing cycles
\item Network packets
\end{itemize}
\textbf{Entry Points.} Given the challenges of securing clusters we've
discussed in section \ref{sec:Security-Challenges-of}, there are
many entry points which the attacker might utilize to compromise the
cluster. These include, 

\begin{itemize}
\item Known vulnerabilities in SSH
\item Remote cluster management software
\item Open ports
\item Stolen login information
\item Rogue processes/rootkits
\end{itemize}
We now present the CIA threat model for cluster security. To do so,
we describe how the enumerated entry points can be used in order to
gain access to cluster resources in order to launch attacks against
Confidentiality, Integrity and Availability of the cluster.

\subsection{CIA Model}

The Confidentiality-Integrity-Availability model is an attractive
way to differentiate attacks. The three properties of the model are
key aspects that must be guaranteed in a secure computing system.
Although there may be some overlap in how an attack can be categorized
(for example, a confidentiality attack can also be an intergrity attack),
we choose to group attacks to a single group explicitly

\subsubsection{Confidentiality}

Confidentiality ensures that only the entities autorized to read information
and access resources can do so. Confidentiality attacks focus on gaining
access to resources without having the proper authorization to do
so\cite{key-1}. Gaining illegitimate root access to the system represents
the ultimate confidentiality attack, however, an attacker may still
learn much without the capability to login to the cluster.

\begin{itemize}
\item \emph{Snooping on External Network.} Cluster users are frequently
allowed to submit jobs over the network. This presents a number of
snooping opportunities. An attacker may learn when a certain user
is running a job (and on which node, if the user is allowed to connect
to compute nodes directly). An attacker may attempt to capture user
data being transferred onto the cluster, or correlate input/output
transmissions to infer information about the size of the job the user
is running.
\item \emph{Snooping on Internal Network.} Messages on the internal cluster
network are often left unencrypted for efficiency reasons. If a malicious
insider attacker manages to gain access to the communication fabric,
or by capturing a node and putting it in promiscuous mode, an attacker
can easily intercept data and control packets.
\item \emph{Scheduler/Metadata Compromise.} If an attacker manages to get
administrative access to the head node of the cluster, he will be
able to examine scheduler logs and job metadata information to learn
about currently running jobs and what jobs have users run in the past. 
\item \emph{Resource Subversion - Computational.} An attacker who has gained
access to a single cluster node, may, unless the cluster is specifically
configured to disallow this, bypass the scheduler altogether and launch
a job from shell. This grants the attacker unauthorized access to
the computational resources of the cluster, enabling him to perform
illegitimate tasks such as hashing values for an offline dictionary
attack.
\item \emph{Resource Subversion - Storage.} Likewise, an attacker who gained
access to a single cluster node may implicitly also gain access to
the storage resources of the cluster. The storage may be used to house
and serve warez, pornography or other illegal material. This is especially
severe if the attacker gains enough access to open an unprotected
port on the cluster through which such connections can be handled.
\end{itemize}

\subsubsection{Integrity}

Integrity ensures that all modifications done to the data are done
by entities that are authorized to do so. Integrity attacks violate
this condition by enabling modification for entities not approved
for doing so. Modification is understood to mean creating, changing,
appending, writing and deleting user and meta data.

\begin{itemize}
\item \emph{Internal Network Packet Injection}. An attacker who gains access
to the internal network of cluster can use it to send legitimate-looking
packets with incorrect data to other nodes. For example, the attacker
may attempt to subvert a computation by sending packets with incorrect
data in them.
\item \emph{Scheduler Tampering}. An attacker who has gained administrative
access to the scheduler may tamper with it, in order to preempt other
jobs running on the cluster (this can also be classified as an availability
attack), or to give his own job higher priority.
\item \emph{Log Tampering}. Cooperative clusters often allocate a quota
of computing cycles for each user. By tampering with logs, an attacker
can modify other people's remaining quotas or his own quota, if the
attacker is a legitimate user.
\item \emph{Data Tampering}. With a significantly authorative access, the
attacker can modify user data on the storage nodes at a whim. In lieu
of sufficient backup, this can be particularly disasterous, since
data residing on the storage nodes is generally the result of hundreds
of hours of computations.
\end{itemize}

\subsubsection{Availability}

Availability ensures that the resources are available to people who
are authorized to access them when they wish to access them. The goal
of an availability attack is to make the resource unavailable to the
intended users - what is knows as a \emph{Denial of Service} attack.
Unlike Confidentiality and Integrity, which have been extensively
studied and modeled, availability is a more fleeting property to define. 

\begin{itemize}
\item \emph{Exhausting Log Space}. Depending on the cluster configuration,
exhausting log space may be an effective availability attack if the
cluster is configured to reject any submitted jobs that it is unable
to log. 
\item \emph{Exhausting Scratch} \emph{Space}. An impropertly configured
cluster may allow users to store data on the same partition that is
used for computation scratch space by the cluster. Exhausting this
space will cause the cluster to have insufficient disk storage to
execute a job. 
\item \emph{Exhausting Storage Space}. An attacker may attempt to compeletely
fill up the existing storage space, making it impossible for legitimate
users to store the results of their jobs.
\item \emph{Scheduling DoS}. An attacker may schedule a repetetive, non-expiring
job (such as a simple program with an infinite loop) to run on all
the computing nodes of the cluster, thus denying legitimate users
the ability to launch jobs.
\end{itemize}

\section{Summary and Conclusions\label{sec:Summary-and-Conclusions}}

In this paper, we have presented a threat model for cluster computing.
Computing clusters differ from traditional networks in design and
approach, and present a unique security challenge compared to them.
Additionally, a cluster has several properties which make it a highly
desirable target for an attack. In their current state, many clusters
have a shell of varying crunchiness on the outside, and a very soft,
unprotected inside. We presented a CIA model which demonstrates that
upon breaking into a cluster through a single node, there is very
little limit imposed on what the attacker can do while inside, in
terms of Confidentiality, Integrity and Availability. The inherent
conclusion that can be drawn is that while securing the individual
nodes and preventing break-ins is important, much effort need to be
put into securing the cluster from the inside out, with the emergent
properties of cluster computing in mind.


\begin{thebibliography}{10}
\bibitem{baker}Mark Baker, Geoffrey Fox, Hon Yau. {}``Cluster Computing Review''.
Center for Research on Parallel Computing. Tech. Rep. CRPC-TR95. 1995.
\bibitem{baker2}Mark Baker. {}``Cluster Computing White Paper''. University of Portsmouth.
V. 1.01. 2001. 
\bibitem{rahib}Rahib Hasan, Suvda Myagmar, Adam J. Lee, William Yurcik. {}``Toward
a threat Model for Storage Security''. International Workshop on
Storage Security and Survivability (StorageSS) in conjunction with
12th ACM Conference on Computer and Communications Security (CCS 2005)
, November 11, 2005.
\bibitem{krebs}Brian Krebs. {}``Hackers Strike Advanced Computing Networks''. Washington
Post. 04/13/2004.
\bibitem{microsoft-threadmodel}Microsoft Corp. {}``Threat Model Your Security Risks''. MSDN Magazine.
Nov. 2003. 
\bibitem{suvda}Suvda Myagmar, Adam Lee, William Yurcik. {}``Threat Modeling as a
Basis for Security Requirements''. Symposium on Requirements Engineering
in Information Security, August 2005.
\bibitem{naqvi}Syed Naqvi, Michel Riguidel. {}``Threat Model for Grid Security Systems''.
European Grid Computing Conference 2005 (EGC2005).
\bibitem{perrine-teracrack}Tom Perrine, Devin Kowatch. {}``Teracrack: Password cracking using
TeraFLOP and PetaByte Resources''. San Diego Supercomputer Center,
2003. http://security.sdsc.edu/publications/teracrack.pdf
\bibitem{key-1}Charles Pfleeger, Shari Pfleeger. \emph{Security in Computing}. Prentice
Hall, 2003.
\bibitem{Pourzandi-"clustersec"}Makan Pourzandi, David Gordon, William Yurcik, and Gregory A. Koenig,
\char`\"{}Clusters and Security: Toward Distributed Security for Distributed
Systems ,\char`\"{} IEEE Cluster Computing and Grid (CCGrid) , May
2005. 
\bibitem{microsoft}Frank Swiderski, Window Snyder, \emph{Threat Modeling}, Microsoft
Press, 2004.
\bibitem{top500}Top 500. {}``Charts for June 2005 - Architecturs''. Available at
http://www.top500.org/lists/2005/06/charts.php?c=1.
\bibitem{top500-ibm}Top 500. {}``System Info: BlueGene/L''. Available at http://www.top500.org/sublist/System.php?id=7605.
\bibitem{torres}Miguel Torres, Rayford B. Vaughn, German Florez, Zhen Liu, Susan M.
Bridge, {}``Attacking a High Performance Computer Cluster'', Proceedings
of the 15th Annual Canadian Information Technology Security Symposium,
May 2003.
\bibitem{Yurcik1-EmergentProperties}William Yurcik, Gregory A. Koenig, Xin Meng, and Joseph Greenseid,
\char`\"{} Cluster Security as a Unique Problem with Emergent Properties:
Issues and Techniques ,\char`\"{} 5th LCI International Conference
on Linux Clusters , Presentation, May 2004.\end{thebibliography}
\end{document}